\documentclass[12pt]{article}
\usepackage{epsf,amsfonts,amsbsy}
\newcommand{\be}{\begin{equation}}
\newcommand{\ee}{\end{equation}}
\newcommand{\YM}{Yang-Mills}
\newcommand{\mh}{{\bf \hat{\mu}}}

\newcommand{\nh}{{\bf \hat{\nu}}}
\newcommand{\rh}{{\bf \hat{\rho}}}
\newcommand{\Amu}{{\bf A}_{\mu}(x)}
\newcommand{\Anu}{{\bf A}_{\nu}(x)}
\newcommand{\Amua}{A_{\mu}^a(x)}
\newcommand{\Amal}{\mathbf{A}_{\mu}^{(\alpha)}(x)}
\newcommand{\Amub}{\mathbf{A}_{\mu}^{(\beta)}(x)}
\newcommand{\LAa}{\boldsymbol{\lambda}_a}
\newcommand{\Lada}{\boldsymbol{\lambda}_a^{(adjoint)}}
\newcommand{\LAb}{\boldsymbol{\lambda}_b}
\newcommand{\LAc}{\boldsymbol{\lambda}_c}
\newcommand{\Fmunu}{{\bf F}_{\mu \nu}(x)}
\newcommand{\Ftmunu}{\tilde{\bf F}_{\mu \nu}(x)}
\newcommand{\Fpmunu}{{\bf F}'_{\mu \nu}(x)}
\newcommand{\Ftpmunu}{\tilde{\bf F}'_{\mu \nu}(x)}
\newcommand{\Dmu}{\partial_{\mu}}
\newcommand{\Dnu}{\partial_{\nu}}
\newcommand{\BPL}{\boldsymbol{[}}
\newcommand{\BPR}{\boldsymbol{]}}
\newcommand{\SUN}{SU(N)}
\newcommand{\UN}{U(N)}
\newcommand{\UO}{U(1)}
\newcommand{\SUT}{SU(2)}
\newcommand{\SUINF}{SU(\infty)}
\newcommand{\ZN}{\mathbf{Z}_N}
\newcommand{\Z}{\mathbf{Z}}
\newcommand{\TF}{T_4}
\newcommand{\RF}{{\bf R}^4}
\newcommand{\EPS}{\boldsymbol{\epsilon}}
\newcommand{\BET}{\boldsymbol{\beta}}
\newcommand{\Om}{\boldsymbol{[}\Omega(x) \boldsymbol{]}}
\newcommand{\Omegamu}{\Omega_{\mu}(x)} 
\newcommand{\Omeganu}{\Omega_{\nu}(x)}
\newcommand{\Omegatnu}{\Omega^+_{\nu}(x)}
\newcommand{\xOnurh}{\Omega_{\nu}(x + \rh) \Omega_{\rho}(x)}
\newcommand{\xOrhonh}{\Omega_{\rho}(x + \nh) \Omega_{\nu}(x)}
\newcommand{\xOnurph}{\Omega'_{\nu}(x + \rh) \Omega'_{\rho}(x)}
\newcommand{\xOrhonph}{\Omega'_{\rho}(x + \nh) \Omega'_{\nu}(x)}

\newcommand{\Onu}{\BPL\Omega_{\nu}(x) \BPR}

\newcommand{\nmunu}{n_{\mu \nu}}
\newcommand{\dmunu}{\varphi_{\mu \nu}}

\newcommand{\Gmu}{\Gamma_{\mu}}
\newcommand{\Gnu}{\Gamma_{\nu}}
\newcommand{\tbc}{twisted boundary conditions}

\newcommand{\eps}{\epsilon_{\mu \nu \rho \sigma}}
\newcommand{\Anm}{\bar{A}_{\nu}^{(\mu)}}
\newcommand{\Arm}{\bar{A}_{\rho}^{(\mu)}}
\newcommand{\Asm}{\bar{A}_{\sigma}^{(\mu)}}
\newcommand{\Anr}{\bar{A}_{\nu}^{(\rho)}}
\newcommand{\kpn}{\kappa(\nmunu)}
\newtheorem{theorem}{Theorem}
\newtheorem{lemma}{Lemma}

\begin{document}

\begin{flushright}
FTUAM-97/18
\end{flushright}
\begin{center}
{\bf Yang-Mills Fields on the 4-dimensional torus.\\ 
Part I: Classical Theory \footnote{This contains the first part of the
lectures  given by the author at the 1997 Advanced School on Non-perturbative
Quantum Physics
(Pe\~niscola)    }}

\vspace{2mm}

A.~Gonz\'alez-Arroyo

\vspace{2mm}

 Depto. F\'{\i}sica Te\'orica C-XI, \\ Universidad Aut\'onoma de Madrid, \\
Cantoblanco, Madrid 28049 , SPAIN\\
\vspace{5mm}
\end{center}

{\bf Abstract}
We   review some of the most important results obtained over the 
years on the study   of Yang-Mills fields on the four 
dimensional torus at the classical level. 

\vspace{2mm}


\newpage
\tableofcontents
\newpage
\section{Introduction}
Yang-Mills fields play a crucial role in our understanding of  Particle Physics. 
Under a relatively simple and elegant formulation, a rich  behaviour hides. 
Phenomena   which are of great importance  in the real world, such as Confinement 
and chiral symmetry breaking, are  supposed to follow from the dynamics of these fields. 
However, these phenomena are highly non-perturbative and still defy ourselves to try to 
derive them properly from first principles.  In trying to address such complicated issues
a good deal of new ideas and strategies have been put forward by different authors over 
the years. Some of these techniques constitute a major breakthrough in our
capacity  to derive quantitative and  qualitative approaches to Quantum
Field Theory (QFT). What in our opinion is the most important new method
 which arose in
this context was the formulation of Yang-Mills theories and other QFTs on a
space-time lattice \cite{Wilson}. This method,   not only gave rise to new
non-perturbative methods of computation, but  also provided a new
illuminating way of understanding  the phenomena taking place in Quantum Field
Theory. Other major contribution followed the work of Polyakov, who 
argued about  the relevance of classical configurations in computing and
explaining some effects in QFT \cite{Polyakov1,Polyakov2}. Most notably the so-called
pseudo-particles, of which the best known example is the {\em instanton}
\cite{BPST}, are known to play an important role in some statistical
mechanical systems and are expected  to do the appropriate in QFT. 

In this review we will study  a different  strategy  put forward by `t
Hooft~\cite{twist1,twist2,twist3}. He considered Yang-Mills fields living in an Euclidean 
 space-time torus $\TF$.  In 't Hooft's original presentation, the interest of
this approach lies in the  characterization of Confinement that it provides, 
and the information it supplies on the possible phases of Yang-Mills theory.  
In very brief terms, we can say that  the topology of the torus gives rise to a
non-trivial topology in the space of Yang-Mills fields which has an
appealing physical interpretation.
The way we see it nowadays, this is just one aspect of the interest that 
a formulation of \YM \ fields on the torus can have. First of all, the torus 
provides a gauge invariant infrared cut-off of the theory.
 For that purpose, we could have taken any other Riemannian 4-dimensional 
compact manifold, but the torus has several advantages which we will now list:
\begin{itemize}
\item  The group of traslations is abelian, and hence one might still
use Fourier decompositions. Furthermore, one can use a flat metric in all
points of space-time.
\item One can regard all configurations on  the torus as configurations in 
$\RF$ which are periodic. In other words, one can glue up configurations
 defined on the torus, to make up configurations on $\RF$.  
\item In the lattice formulation of gauge theories, numerical methods demand a
finite number of degrees of freedom. Henceforth, one is forced to
put the system in a finite volume.  For reasons similar to the ones mentioned
before, one usually imposes  periodic boundary conditions, which is equivalent
to having the system living on the torus. The study of Yang-Mills on the
torus allows us to understand some  of the observed  finite size effects.
One can  take the continuum limit, in  a way such that the 
  size of the torus stays finite in physical units. 
  \item Topology of the torus is non-trivial (both $\pi_1(\TF)$ and
$H_i(\TF)$ are non-zero). This allows the topological characterization 
of some properties of the theory, including Confinement. This is basically 
't Hooft's original motivation. In addition, as is the case in other theories,
boundary conditions  serve to stabilize several classical configurations.
\item A curious phenomenon takes place in the large N limit. 
Eguchi and Kawai \cite{ek} argued, that in that limit,  some 
quantities might have 
 no finite-size effects. Hence, the torus would 
give results which are identical to those of $\RF$.  
\item The size of the torus provides a parameter, which can be used to 
interpolate between the perturbative (small size) region and the 
non-perturbative (big size) one. This can provide a method for approximately 
calculating some non-perturbative quantities, as argued by L\"uscher~\cite{luscher1}. 
In addition, if the ideas of the author are correct, this can provide an 
insight into the structure of the Yang-Mills vacuum and the origin of
 Confinement  \cite{Investigating}.

\end{itemize}
To conclude, we can say that the     study of \YM \ on the 4-D torus serves as
 an example of the   behaviour of the fields on other  non-simply
 connected  manifolds. The latter can be relevant in other contexts.

In these lectures, we will  review the most important  results obtained by 
different authors on this topic.  They  can be split into two parts:
classical and quantum theory. In this paper we will present the first part. 
The quantum  part, which will follow, will make use of the results 
presented here. 

Our aim is to make a pedagogical and self-contained introduction to the 
subject. Accordingly, we will   present all the ideas and results in the 
most simple language. We will nevertheless frequently comment upon the 
more precise mathematical statement of
those results. In other cases, we will refer the readers to the relevant
publications.

 The layout of the paper is shown in the Contents.

\section{Notation and general formulas}
Let us consider a 4-dimensional torus $T_4$ of size $l_0 \times l_1 \times
l_2 \times l_3$. The metric is Euclidean and we take coordinates $x$
($x_{\mu} $ for $ \mu = 0 \ldots 3$) on the torus as follows:
\begin{equation} 
0 \le x_{\mu} <  l_{\mu}\ .
\end{equation}
Let us introduce the symbol $\mh$ $\equiv (0, \dots,l_{\mu},0, \dots) $
which has all but the  $\mu$th  component equal to zero.

We now consider \YM \ fields  defined on $\TF$. For simplicity we will
 restrict ourselves to the $\SUN$ gauge groups. We will express the vector 
potential fields in  the usual  
matrix form:
\be
 \Amu = \Amua  \ \LAa\ ,  
\ee
where $\LAa$ are the generators of the $\SUN$ Lie algebra in the fundamental 
representation, normalized by:
\begin{eqnarray} 
Tr ( \LAa \LAb ) &=& \frac{\delta_{a b} }{2 } \nonumber \\*
  {[} \LAa, \LAb {]} &=& \imath f_{a b c } \LAc\ .
\end{eqnarray}

The expression for the field tensor $\Fmunu$ in terms of the vector potential $\Amu$
is given by: 
\be 
 \Fmunu = \Dmu \Anu -\Dnu \Amu - \imath\, [ \Amu , \Anu]\ .
\ee
Gauge transformations operate on the vector potential $\Amu$ as follows:
\be
\label{gaugetrans}
\Om \Amu= \Omega(x) \Amu \Omega^{+}(x) + \imath\, \Omega(x) \,\Dmu
\Omega^{+}(x)\ .
\ee

In a more  precise mathematical language, \YM\  fields on the torus
are connections on a principal fiber bundle whose base space is $\TF$.
 A good introduction to this formulation
 aimed to  an audience of physicists can be found in Ref. \cite{egh}.
In order to formulate
the theory,  one has first to define an $\SUN$ principal fiber bundle over $\TF$.
This can be done by giving a covering of the torus by open contractible sets
$U_{\alpha}$. These sets can be considered gauge-coordinate patches. In the
overlapping regions ( $x \in U_{\alpha}\cap U_{\beta}$ ), the consistency of
the two descriptions implies that the two gauge-coordinates are  related by
an SU(N) gauge transformation $\Omega_{\alpha \beta}(x)$, which is called a
transition function. The open sets and 
the transition functions define the principal bundle.

Connections define a way to parallel transport from one point of space to
another along some path. Within one  patch $U_{\alpha}$ the parallel 
transport matrices are given by the well-known  ordered exponential formulas in
terms of the vector potentials $\Amal$ for this patch:
\be
 U(\gamma:P \rightarrow Q) = T\exp{\left( -\imath \int_{\gamma} \Amal
dx_{\mu}\right) } \ ,
\ee
where ordering within the path goes from left to right. In the overlapping
regions between  two patches  ( $x \in U_{\alpha}\cap U_{\beta}$ ), one
might use instead the vector potentials $\Amub$, related to the 
previous ones by a gauge transformation involving the transition matrices:
\be
\Amub = \BPL \Omega_{\beta \alpha}(x)  \BPR \Amal\ ,
\ee
which can be considered a change of coordinates for the gauge fields.
The link with  our original  presentation follows by identifying
our vector potential $\Amu$ with that of a  single patch which covers the
whole torus.

\section{Boundary Conditions and Twist}
By continuity one can extend $\Amu$ and $\Fmunu$ to the boundary
of the  hypercube $[0,l_0] \times [0,l_1] \times [0,l_2]  \times [0,l_3]$.
However, one must realize that for a point  $x$   located  on the face
$x_{\nu}=0$, there is a corresponding point  $x + \nh$
which labels    the same point
on the torus.
 Henceforth, the gauge fields ${\bf A}_{\mu}(x)$
and ${\bf A}_{\mu}(x + \nh)$ have to be physically equivalent. For gauge 
fields, equivalence  does not imply   equal vector potentials. It is enough if
they are equal modulo a gauge transformation:
\be
\label{tbc}
{\bf A}_{\mu}(x+\nh) = \Onu {\bf A}_{\mu}(x)\ ,
\ee
where $\Omega_{\nu}(x)$ are elements of the $\SUN$ group which depend only on the transverse coordinates 
$x_{\rho}$ with $\rho \ne \nu$. These matrices are known as {\em twist matrices}. 

If we now consider  a point $x$ located at the two-dimensional surface
$x_{\nu}=x_{\rho}=0$,  consistency of the      previous   boundary conditions demands:
\be
\label{consist}
{\bf A}_{\mu}(\nh + \rh ) =  \BPL \xOnurh \BPR {\bf A}_{\mu}(0) = \BPL
\xOrhonh \BPR  {\bf A}_{\mu}(0)\ . 
\ee
This implies 
\be
\label{twisteq}
\xOrhonh = z_{\rho \nu} \  \xOnurh\ ,
\ee
where the constants $z_{\rho \nu}$ are phases which  can occur  due to the
 quadratic way in which $\Omega$ enters in the gauge transformation formula
Eq.~\ref{gaugetrans}.
Since $\xOrhonh$ and  $\xOnurh$  belong to $\SUN$, the  phases $z_{\rho \nu}$ are elements
of the center of this group ($\ZN$). Hence, we can express
them as 
\be
z_{\mu  \nu} = \exp \{ 2 \pi \imath\, \frac{n_{\mu \nu}}{N} \}\ ,
\ee
where  $\nmunu$ is an  antisymmetric tensor of integers defined modulo N. 
This tensor has 6 independent coefficients, which can be expressed as usual in terms 
of two independent three vectors ($\vec{k}$ and $\vec{m}$) as follows:
\begin{eqnarray}
n_{i j} &=& \epsilon_{i j k } \ m_k \nonumber 
\\
n_{0 i } &=& k_i\ .
\end{eqnarray}
We will refer to these boundary conditions as {\em \tbc}, and to $n_{\mu
\nu}$ as the {\em twist tensor}.
The elements of the  tensor $\nmunu$, being integers, label
topologically inequivalent sectors.
The particular case   $n_{\mu \nu} = 0$ is often referred as {\em no-twist}. 
In particular,  choosing the twist matrices equal to  the identity,
corresponds to this case. The gauge fields are then said to be strictly periodic. 

Under gauge transformations,  the vector potentials transform as 
shown in Eq.~\ref{gaugetrans} , but the twist matrices also change as follows:
\be
\Omega_{\mu}(x) \longrightarrow \Omega'_{\mu}(x)=\Omega(x+ \mh)\,
\Omega_{\mu}(x)\, \Omega^+(x)\ .
\ee
However, notice that the change of twist matrices does not produce any
change of the twist tensor $\nmunu$. 

In addition to gauge transformations, there is another symmetry group present. 
It is the group of transformations:
\be
\label{ZNsym}
\Omega_{\mu}(x) \longrightarrow \Omega'_{\mu}(x) =  z_{\mu} \
\Omega_{\mu}(x)\  ,
\ee
where $z_{\mu}$ are elements of the center. This group is isomorphic to $\ZN^4$
and plays an important role in what follows. 

The previous introduction of \tbc\  follows very closely the one  made by `t~Hooft
in his first papers on the subject \cite{twist1,twist2}. Using the 
language of connections on fiber bundles, the twist matrices are simply 
compositions of different transition functions in the limit when one of the 
patches covers the whole torus~\cite{vanbaal1,vanbaaltesis}. In this
respect, the consistency condition Eq.~\ref{consist} is related to  the
so-called {\em co-cycle} condition, that transition matrices have to satisfy:
\be
\label{cocycle}
\Omega_{\alpha \gamma} = \Omega_{\alpha \beta} \Omega_{\beta \gamma} \, .
\ee
The occurence of the phases $z_{\mu \nu}$ and of the twist tensor, has to do
with the fact that gauge transformations of the vector potentials are
insensitive to the center of the group $\ZN$. Henceforth, one could
consider the gauge group (structure group) as being $\SUN/\ZN$. 
A non-zero twist tensor can be seen as an
{\em obstruction} to go from an  $\SUN/\ZN$ principal fiber bundle
to an $\SUN$ bundle. This obstruction can be related to the second
homology class of the base manifold with coefficients in the center of the
group $H_2(\TF,\ZN)$. For a proof of these facts see for example
Ref.~\cite{sedlacek}.  
\section{Twist Matrices}
The first problem which we will address is the form and existence of
solutions  to the twist matrix equations Eq.~\ref{twisteq}. We will consider 
two classes of  solutions which will be relevant in what follows. The first 
class is given by abelian space-dependent matrices. The second class 
is given by constant non-abelian matrices: {\em twist eaters}. 
\subsection{Abelian twist matrices}

Here we will show that there is  always a solution to the problem of finding twist matrices 
satisfying  Eq.~\ref{twisteq}.  This solution is given in terms of commuting
matrices, and is  hence referred as abelian. 
Let us set:
\begin{equation}
\label{Omabelian}
\Omega_{\mu}(x) = \exp{ \{ \imath\  {\bf \omega}_{\mu}(x) \} }\ , 
\end{equation}
with $[\, {\bf \omega}_{\mu}(x)\, ,\, {\bf \omega}_{\nu}(x) \,] = 0 $. If we diagonalize
the matrices ${\bf \omega}_{\mu}$, the twist condition
Eq.~\ref{twisteq} amounts to the following condition on its eigenvalues
$\omega_{\mu}^a$:
\be
\Delta_{\nu} {\bf \omega}_{\mu}^a(x) - \Delta_{\mu} {\bf \omega}_{\nu}^a(x) = \frac{2
\pi \nmunu }{N}  + 2 \pi q_{\mu \nu}^a\ , 
\ee
where ${\bf Q}_{\mu \nu} = $diag$ (  q_{\mu \nu}^a ) $ is an integer diagonal  matrix
satisfying $Tr ({\bf Q}_{\mu \nu}) = -\nmunu $, and the symbol $ \Delta_{\mu}$ is defined
as:
\be
\label{deltadef}
\Delta_{\mu} \phi(x) = \phi(x + \mh ) -  \phi(x)\ .
\ee
A  particular solution of the previous equation is given by:
\be
\label{partabetm}
{\bf \omega}_{\mu}(x) = \frac{\pi}{N} \nmunu \frac{x_{\nu}}{l_{\nu}} {\bf T}\ ,
\ee
with ${\bf T} = \mbox{diag}( 1, \ldots , 1, 1-N )$. This solution corresponds to
the choice $q_{\mu \nu}^a  = 0 $ for $ a \neq N$ and $q_{\mu \nu}^N  = -\nmunu$.
For arbitrary value  of the integers $q_{\mu \nu}^a$ one gets:
\be 
\label{omegamunu}
{\bf \omega}_{\mu}(x) = \frac{\pi}{N}  \frac{x_{\nu}}{l_{\nu}} {\bf T}_{\mu
\nu}\ ,
\ee
with
\be
\label{Tmunu} 
 {\bf T}_{\mu \nu} = \nmunu {\bf I} + N {\bf Q}_{\mu \nu}\ .
\ee
A general solution can be
obtained by adding to the previous solutions  a term $\Delta_{\mu}  \phi(x)$,
which is a general solution of the
homogenous equation.
This general solution can be easily seen  to be a gauge transformation of 
the solution expressed in Eqs.~\ref{omegamunu}-~\ref{Tmunu}.  
\subsection{Twist eaters}
A class of solutions   which plays an important role in the following
are the so-called {\em twist eaters} or twist eating solutions.
 These are constant matrices $\Omega_{\mu}(x) = \Gmu $ satisfying: 
\be
\label{twisteat}
\Gmu \Gnu = \exp{ \{ \frac{ 2 \pi \imath\, \nmunu} {N} \}}\, \Gnu \Gmu\ .  
\ee

The previous equation defines  a group, whose generators
are $\Gmu$ and the phase $\exp{\{2 \pi \imath / N\}}$, and satisfying
the relations Eq.~\ref{twisteat} involving $\dmunu =  \frac{\nmunu} {N}$. We
will call this group   the {\em twist group} $\cal G$. Notice that the Clifford
Algebra is a particular case of this  class of groups. Our solutions
can be looked at as unitary representations of this group.  Hence, we can use
the standard methods and ideas on representations. We will actually be
 more general that required and work in arbitrary
 space-time dimension $d$ ($\mu=0 \ldots (d-1)$).

Given one solution $\Gmu$ one can construct another one by means of:
\begin{eqnarray}
\label{equivalence}
\Gmu  &\longrightarrow& \Gmu' =  \Omega \Gmu \Omega^+ \\
\label{multipl}
\Gmu  &\longrightarrow& \Gmu' = z_{\mu}  \Gmu\ ,
\end{eqnarray}
with $\Omega \in \SUN $ and $z_{\mu} \in \ZN$.
Furthermore, given two solutions with $\dmunu =\frac{\nmunu^{(1)}}{ N_1} =
\frac{\nmunu^{(2)}}{N_2}$, one can construct the direct sum, which is
an $N_1 + N_2$ dimensional representation. The new representation is reducible. 
One needs only to find the irreducible representations and the rest can be 
formed by direct sum.   

The choice of generators is not unique, and thus different sets of
$\varphi_{\mu \nu}$ correspond to the same  group. To investigate 
this point let us introduce the following notation for the product 
of the generators:
\be 
\label{Gamdef}
\Gamma(n) = \Gamma_0^{n_0}\ \Gamma_1^{n_1}\ \ldots \Gamma_{d-1}^{n_{d-1}}\ ,
\ee
where $n_{\mu}$ are integers.
The matrices $\Gamma(n)$ satisfy:
\be 
\label{Gamprod}
\Gamma(n) \Gamma(m) = \Gamma(m+n)\, \exp \{ \frac{2 \pi i }{N} \sum_{\mu >
\nu} n_{\mu}\,
m_{\nu}\, \nmunu \}\ .
\ee
The matrices of the form $z \Gamma(n)$ form the group $\cal G$. 
One might choose other generators
$\Gmu' = \Gamma( s^{(\mu)} )$ provided the matrix $s^{(\mu)}_{\nu}$  is
invertible. Notice that if we use $\Gmu'$ instead of $\Gmu$, the twist 
tensor changes to:
\be 
\nmunu \longrightarrow \nmunu' = \Sigma s^{(\mu)}_{\rho}  s^{(\nu)}_{\sigma}
n_{\rho \sigma}\ .
\ee
For irreducible representations, by Schur lemma,  all the matrices which
conmute with the generators (elements of the center)  must be multiples of
the identity. In particular,
any matrix of the group taken to the Nth power belongs to the center. This
allows the numbers $s^{(\mu)}_{\nu}$ to be considered integers $\bmod N$.
This enlarges the group of transformations of generators from $SL(d,\Z)$ to 
$GL(d,\ZN)$. 
One can use these transformations to bring $\dmunu$ to a canonical form:
\be
\dmunu(\mbox{\small canonical})= \left( \begin{array}{cc} \begin{array}{cc} 0 &
{\bf \Delta} \\ - {\bf \Delta} & 0  \end{array} & 0 \\
0 & \begin{array}{cc}\ddots & 0\\0 &0 \end{array}   \end{array}  \right)\ ,
\ee
where ${\bf \Delta} =\ $diag$(\frac{1}{p_1}, \ldots, \frac{1}{p_r})$. The
integers $p_i > 1$ are such that $p_i$ divides  $p_j$ for $ i > j $.
The value of $r \le [\frac{d}{2}]$ will be referred to as the {\em rank}
of the group $\cal G $. An
important subgroup is the center $\cal C$ of $\cal G $. It is generated by
$(\Gamma'_{i-1})^{p_i}, (\Gamma'_{r+i-1})^{p_i} $ for $ i = 1 \ldots r$ and $\Gamma'_{2r}
\ldots \Gamma'_{d-1}$. For irreducible representations, all these matrices 
have to be a multiple of the identity $\tau_{\mu} \bf I$, and the constants
$\tau_{\mu}$ characterize the representation in question. 

Now let us use the
standard strategy of considering the maximal abelian subgroup $\cal A $. It
is generated by $\Gamma'_{i-1}\, ,\, i=1 \ldots r$ (factoring out the center).
Let
$ |\vec{\lambda} >$
denote a vector which is a simultaneous
eigenstate of  these matrices with eigenvalues $\lambda_{i-1}$ (which can
be written as an r dimensional weight vector $\vec{\lambda}$). The matrices
$\Gamma'_{r+k-1}$ now act on this state and transform it  into another 
eigenstate with eigenvalue $ (t_k \vec{\lambda})_i
\equiv \lambda_{i-1}\,  \exp \{\frac{2 \pi \imath}{p_i}
\delta_{i k} \}$.
In this way,  given one state $ |\vec{\lambda}_0 >$ one generates $p_1\,p_2\
\cdots p_r$ different states $ |\vec{\lambda} >$ ( Notice that
$\lambda_{i-1} \ne 0$ since it is a representation of a group, and the matrices
have to be invertible).
Now we can normalize these eigenstates  so  that:   
\be
          \Gamma'_{r+k-1} |\vec{\lambda} > = \chi_k  |t_k \vec{\lambda} > \ .
\ee
The definition is consistent because  the $\Gamma'_{r+k-1}$ with
different $k$ conmute. We have to insert $\chi_k$ because
$\chi_k^{p_k}=\tau_{r+k-1}$
is a Casimir, which is not necessarily equal to 1.

Thus, we have explicitly constructed the irreducible
representations of the twist group. The dimension of the representation
is:
\be
N_0= p_1 \times p_2 \times \cdots p_r\ .
\ee
Henceforth, supposing that there is a solution in $N \times N$ matrices,
then the representation is irreducible if and only if $N = N_0$.
One can construct  reducible solutions in all dimensions which
are multiples of $N_0$. 

The irreducible representation is unique modulo similarity transformations
(Eq. ~\ref{equivalence}) and multiplication of the matrices by a
constant (Eq. ~\ref{multipl}).
Representations that
are related by a similarity transformation are called equivalent.
As we can see, equivalent representations
are labelled by the Casimirs:  $\tau_{i-1}=\lambda_{i-1}^{p_i}$,
$\tau_{r+i-1}= \chi_i^{p_i}$ (for $i =
1 \ldots r$) and $\tau_{\mu}$ (for $\mu = 2r, \ldots , d-1$). For  unitary
representations  these Casimirs are  phases. We are mainly interested
in representations in terms of $\SUN$ matrices. In that case, the
phases have a discrete set of values ($ \in \ZN$) and the number of inequivalent
irreducible representations is discrete. We will now calculate 
how many of them there are, since this will turn out to be relevant later on.
We get:
\be
\mbox{\# of irr. reps.} = \left(\frac{N_0}{p_1}\right)^2 \ \cdots
\left(\frac{N_0}{p_r}\right)^2 \dot
N_0^{d-2r} \ = N_0^{d-2}\ .
\ee
This follows from counting the possible values of the Casimirs consistent
with the requirement that the matrices belong to  $\SUN$.
Having studied the problem in general, let us now take a brief look at the 
 2, 3 and 4-dimensional cases, which are the interesting ones for us.

In  2 dimensions, we have 2 matrices $\Gamma_0$ and $\Gamma_1$ and a single
twist tensor element $n_{0 1 } = \varphi_{0 1} N$. If $q=\gcd(n_{0 1 },N)$
then there $\exists$ a couple of integers $n_1$ and $n_2$ such that:
\be
 n_1 N + n_2 n_{0 1} = q\ . 
\ee
Now the canonical generators are $\Gamma_0'=\Gamma_0^{n_2}$,
$\Gamma_1'=\Gamma_1$, and $p_1=\frac{N}{q}$. The representation is
irreducible  if $p_1=N$ ($\Rightarrow q = 1$). Our canonical
basis amounts to $\Gamma_0'=\mathbf{Q}_N$ and $\Gamma_1'=\mathbf{P}_N^+$
where:
\begin{eqnarray} 
   (\mathbf{P}_N)_{k j}  = & z \ \delta_{j\, k+1}   \nonumber
   \\
   (\mathbf{Q}_N)_{k j}  = & z \ \delta_{j\, k}  \exp\{\frac{2 \pi i\,
k}{N}\}\ , 
\end{eqnarray}
where indices are defined modulo N, and z is chosen  to ensure
determinant equal to unity (z=1 for odd N and $\exp\{\frac{ \pi i }{N}\}$ for even
N). Thus, the solution to our problem is $\Gamma_0=\mathbf{Q}_N^{n_{0 1}}$ and
$\Gamma_1=\mathbf{P}_N^+$, which is unique modulo similarity transformations 
(if N and $n_{0 1}$ are coprime).

In three dimensions we have $n_{i j} = \epsilon_{i j k } m_k$. Then the
3 canonical matrices are $\Gamma'_1 = \Gamma(\vec{s}^{(1)}) ,\ \Gamma'_2 =
\Gamma(\vec{s}^{(2)}) \mbox{and} \  \Gamma'_3=\Gamma(\vec{m}/m_0)$, where
$m_0=\gcd(\vec{m},N)$. The vector $\vec{s}^{(1)}$ can be chosen among those
that satisfy: $ \gcd(\vec{s}^{(1)}\times \vec{m},N)=m_0$. The other vector
is such that $(\vec{s}^{(1)}\times \vec{m}) \cdot \vec{s}^{(2)}=m_0 \bmod N$.
One can easily see that $p_1=N/m_0$ and the representation is irreducible
for $m_0=1$. In this case the generator of the center of the group
must be a multiple of the identity: $ \Gamma'_3=z_3 \mathbf{I}$. The
inequivalent irreducible representations are labelled by $z_3$. 

Finally, we arrive at the 4-dimensional case. We know that there are 2
subcases: rank 1 and rank 2 (rank 0 is trivial). The relevant quantities
are $p_1$ and $p_2$ (for rank 2). In order to obtain these numbers it is not
necessary to bring $\nmunu$ ($\dmunu$) to the canonical form. They can be determined in
terms of invariants of the transformations. We see that $q_1=\gcd(\nmunu,N)$
is an invariant. By looking at the canonical form, one sees that:
\be
p_1=\frac{N}{\gcd(\nmunu,N)}\ .
\ee
If we perform  $SL(4,\Z)$ transformations to $\nmunu$, there is
another invariant:
\be
\label{kpndef}
\kpn \equiv  \frac{1}{4} \nmunu \tilde{n}_{\mu \nu} = \vec{k} \cdot \vec{m}
= P\!f(\nmunu) \ , 
\ee
where $\tilde{n}_{\mu \nu}=(1/2)\, \eps\,  n_{\rho \sigma}$ and $P\!f(\nmunu)$
stands for the Pfaffian of the antisymmetric matrix $\nmunu$. However, $\kpn$
is non-invariant under the wider group of transformations
$GL(4,\ZN)$, or under changes of $\nmunu$ by integer multiples of
N. By first bringing $\nmunu$ to a canonical form with $SL(4,\Z)$
transformations and later applying the wider class of transformations, it
is not hard to see that the other invariant $p_2$ can be obtained as follows:
\be
p_2=\frac{N}{\gcd(\frac{\kpn}{q_1},N)} \equiv  \frac{N}{q_2}\ .
\ee
We remind you that     the necessary and sufficient condition for the
existence of solutions is that $N_0 \equiv p_1 p_2$ divides $N$. The representation
is irreducible for $N_0=N$. The case of rank 1 corresponds to $p_2=1
\Leftrightarrow
q_2=N$. One can summarise our results in the  following statements:
\begin{itemize}
\item The necessary and sufficient condition for the existence of solutions
is that $\gcd(\kpn,N)=N$. This case is called {\bf Orthogonal twist},
since it corresponds to $ \vec{k}\cdot \vec{m} = 0 \bmod N$. The opposite case
is refered to as  {\bf Non-Orthogonal twist}.
\item The representation is irreducible, and hence unique modulo the
transformations~\ref{multipl}-\ref{equivalence}, provided $\gcd(N,\nmunu,\kpn/N)=1$.
It must be one of the $N^2$ inequivalent irreducible representations.
\item The representation has rank 2 provided $\gcd(\frac{\kpn}{q_1},N) \ne N$ 
\end{itemize}

Let us conclude this section by pointing to the relevant references where
the study of twist eating solutions was addressed.   
The first solutions of this type were found in Ref \cite{gjk-a} and \cite{af}.
`t Hooft addressed and solved the problem for the case of rank=1 in 4
dimensions~\cite{stsds}. This also covers the 2 and 3 dimensional cases, the latter
studied previously in Ref. \cite{af}. Rank 2 solutions, whose existence was
established in Ref.~\cite{vanbaal1}, appear and become relevant in the
Twisted-Eguchi-Kawai model \cite{tek1,tek2}. This triggered renewed interest
in finding a complete solution to the problem. The final solution up to 
4 dimensions was obtained in Refs.~\cite{vanbaal2,brihaye}. Finally, the 
d-dimensional case was studied in Refs.~\cite{vbvg,lpoli}.

\subsection{The algebra of twist-eaters}

In this subsection we would like to consider the case of irreducible twists.
Let us examine here what would be the equivalent of the Clifford algebra
for our case. We consider  the $\Gamma(n)$ matrices defined in
Eq.~\ref{Gamdef}. We will see in what follows
that there are as many linearly  independent such matrices as there are elements
in the  quotient group $\cal G/C$. This group is abelian and isomorphic
to $\Z^{p_1} \times \Z^{p_1} \times \ldots \times \Z^{p_r} \times \Z^{p_r}$. 
As a consequence of our study
of the last subsection we  know  that the order of this group is  $N^2$.

Since the representation is irreducible the matrices corresponding to an
element of the center group $\cal C$ are multiples of the identity.
By virtue of the multiplication formula Eq.~\ref{Gamprod}, the matrices
corresponding to two  elements belonging to the same class in $\cal G/C$,
differ by multiplication by a phase. Hence, at most there are as many
linearly independent matrices as elements in $\cal G/C$. Now let us prove
the following result:
\begin{theorem}
If for each class $a \in {\cal G/C}$ we choose a representative: 
\be
\label{lambdef}
\lambda(a) = e^{\imath \phi(a)}\, \Gamma(n(a)) \ , 
\ee
the corresponding matrices are 
linearly independent. 
\end{theorem}

The proof is very similar to the corresponding one for the Clifford algebra.
One makes use of the following result:
\begin{lemma}
Let  $\mathbf{A}$ and $\mathbf{B}$ be two invertible  $N \times N$ matrices  such that
$ \mathbf{A}\, \mathbf{B} = z\,  \mathbf{B}\, \mathbf{A}$ with $z
\ne 1 $, then $z
\in \ZN$ and $Tr( \mathbf{A})= Tr( \mathbf{B}) = 0$.    
\end{lemma}
This lemma can be easily proven. By diagonalizing $A$, for example,  one
reaches  the condition $ (A_{i i}-z\,A_{j j})\,B_{i j} =0$. Now since
$\mathbf{A}$ is invertible $A_{i i} \ne 0$, and hence $B_{i i } = 0$.
This shows that indeed the trace of $\mathbf{B}$ is zero. Now, since
$\mathbf{B}$ is invertible, for any $i$ there must exist $i'$ such that
$B_{i i'} \ne 0$. Hence, for this pair we must have $A_{i i}-z\,A_{i' i'} = 0 $.
If we start with $i'$ and repeat the argument, there must exist another $i''$
satisfying the same relation. Since the number of different eigenvalues of
$\mathbf{A}$ is finite, this implies that there is an integer $p$ such
that $z^{-p} =1$. Hence starting from one eigenvalue one generates a sequence
of $p$ different ones. Now suppose $\mathbf{A}$ has one eigenvalue
$\tau$ with degeneracy $s$. Then  as we have seen $z^{-1}\, \tau$ must also be an
eigenvalue. In addition, its degeneracy must be  $s$ as well. This is so because
$\mathbf{B}$ induces an homomorphism between the spaces  of both
eigenvalues, and due to the invertibility of $\mathbf{B}$, it must be an
isomorphism. In this way, from each eigenvalue one generates $p s$ vectors.
Repeating the operation for an eigenvalue not contained in this set, we
conclude $N=p\,(\sum_i s_i)$, and hence $p$ divides $N$. By diagonalising
$\mathbf{B}$    instead, one can show that the trace of $\mathbf{A}$ is also
zero.
This completes the proof of the lemma.
In fact, our lemma is very much related to  the explicit construction of
irreducible representations of $\cal G$ done in the previous subsection.

As a consequence, the trace of all $\Gamma(n)$  matrices is zero, except for
those which are elements of the center group. Now let us proceed to
prove the  linear
independence, asserted in the Theorem.  We will  work by {\em reductio ad adsurdum}.
Consider that there
exist a  non-trivial linear combination of the $\lambda(a)$ matrices  which is zero:
\begin{equation}
 \sum_{a \in {\cal G/C}} c(a)\, \lambda(a)\ =0\ . 
\end{equation}
Then, if we multiply both sides of the equation  by $\Gamma^{-1 }(n(a))$ and take the
trace, we obtain $N\, e^{\imath \phi(a)} c_a =0$. Since this is true for all $a$, all the
coefficients must be zero, thus  contradicting  the non-triviality
assumption.\ {\em QED}    

Our previous theorem impies that 
the $\lambda(a)$ matrices define a
basis of the complex vector space of all $N \times N$ matrices.
The product
of these matrices satisfies:
\be
\lambda(a)\, \lambda(b)\,=\, e^{\imath d(a,b)} \lambda(a+b)\ , 
\ee
where, by virtue of Eq.~\ref{Gamprod}, $d(a,b)$ is given by:
\be
d(a,b)=\phi(a)+\phi(b)-\phi(a+b) +\sum_{\mu > \nu} n_{\mu \nu}\, n_{\mu}(a)\,
n_{\nu}(b)\  \bmod 2 \pi\ .
\ee
The phases $d(a,b)$ depend upon the choice of representative
($\phi(a),n(a)$). However, the antisymmetric combination:
\be
<a,b> \equiv d(a,b)-d(b,a) = \sum_{\mu \nu} n_{\mu \nu}\, n_{\mu}(a)\,
n_{\nu}(b)\  \bmod 2 \pi  
\ee
does not depend on the choice.

The real subspace of hermitian $N \times N$ matrices is isomorphic to the Lie
Algebra of $\UN$. This  space can be obtained  by linear combinations of the
$\lambda(a)$ matrices with  coefficients  satisfying certain relations. Using
the freedom to make  specific choices of the representatives
($\phi(a),n(a)$), one can simplify the form of the relations.
For example, by  choosing the  matrices, such that
$\lambda(-a)=\lambda^+(a)$, the condition of hermiticity reads  $c(-a)=c^*(a)$.
Hence, one can obtain the form of the structure constants of U(N) in the
basis of   $\lambda(a)$ matrices, as follows:
\be
f_{a b c}=   - \imath\, \delta_{c\,a+b}\, (e^{\imath d(a,b)} - e^{\imath
d(b,a)}) \ .
\ee
 Other choices of representatives can impose  certain properties on the
$\lambda(a)$ matrices.

The algebra of twist-eaters was introduced  in Refs.~\cite{en,tek2}
in 2 and 4 dimensions respectively. In those references, it was used as a 
basis of the Lie Algebra of $\SUN$. This  will turn out to be useful later (See also
Refs.~\cite{fk-a,lusc}).

\section{Topology of gauge fields}
In addition to twist, we must consider the ordinary
{\em topological charge} (instanton number):
\be
Q=\frac{1}{16 \pi^2}\int_{T_4} Tr(  \Fmunu\, \Ftmunu)\, d^4x\ .
\ee
Using the fact that $Tr( \Fmunu\, \Ftmunu)$ is a pure divergence
($\partial_{\mu} K^{\mu} $), and integrating once, we get:
\be 
Q= \sum_{\mu} \int \frac{d \sigma_{\mu}}{16\ \pi^2 } \ \Delta_{\mu} K^{\mu}\ ,
\ee
where  $\Delta_{\mu}$ is defined in Eq.~\ref{deltadef},  $d \sigma_{\mu}$ is the
integral over the 3 dimensional face $x_{\mu}=0$,
 and $ K^{\mu} $ is given by:
\be
 K^{\mu} = 2\, \eps\,  Tr( A_{\nu} \partial_{\rho} A_{\sigma}
-\frac{2}{3}\imath\, A_{\nu} A_{\rho} A_{\sigma})\ .
\ee
The main ingredient that we need now, is how does $K^{\mu}$ change under
a gauge transformation. We have: 
\be
\label{Kgt}
K^{\mu}(\BPL \Omega \BPR A) = K^{\mu}(A) -\frac{2}{3}\imath\, \eps\, Tr(\bar{A_{\nu}}
\bar{A_{\rho}} \bar{A_{\sigma}})  -2\, \eps\, \partial_{\rho} Tr (
\bar{A_{\nu}} A_{\sigma})\ ,
\ee
with $\bar{A_{\nu}} = \imath\, \Omega^+ \partial_{\nu}\Omega$.
In our case $A_{\nu}(\mh) = \BPL \Omega_{\mu}  \BPR  A_{\nu}(0)$, and hence we get:
\begin{eqnarray}
Q=& -\frac{\imath}{24 \pi^2}\, \eps \int d \sigma_{\mu}\,  Tr(  \Anm \Arm
\Asm ) \\
&- \frac{1}{8 \pi^2}\, \eps \int d \sigma_{\mu \rho }\, Tr ( \Delta_{\rho} (\Anm
A_{\sigma}) )\ , 
\nonumber
\end{eqnarray}
with $ d \sigma_{\mu \rho }$ the  integral over the 2-dimensional surface
$x_{\mu}=x_{\rho}=0$, and
$\Anm= \imath\, \Omega_{\mu}^+ \partial_{\nu}\Omega_{\mu}$.
Now using $D_{\rho} \Anm = D_{\mu} \Anr$, with $D_{\rho}$ defined by
\[
D_{\rho} \Anm =\Omega_{\rho}^+(\Delta_{\rho} \Anm +\Anm) \Omega_{\rho}
-\Anm\ ,
\]
and manipulating the previous equation, we arrive at:
\begin{eqnarray} 
\label{topch}
Q=&-\frac{1}{24 \pi^2}\, \eps \int d \sigma_{\mu}\,  Tr( \Omega_{\mu}^+ (\partial_{\nu} \Omega_{\mu})
\Omega_{\mu}^+ (\partial_{\rho} \Omega_{\mu}) \Omega_{\mu}^+
(\partial_{\sigma} \Omega_{\mu})) \\
&-\frac{\imath}{8 \pi^2}\, \eps \int d \sigma_{\mu \rho }\, Tr( (\Delta_{\rho} \Anm +
\Anm  )\, \Omega_{\rho} \partial_{\sigma} \Omega_{\rho}^+)\ ,
\nonumber
\end{eqnarray}
which is our final formula. As expected, only the  twist matrices $\Omega_{\mu}$ 
enter the formula.
Hence, given the twist   matrices, one can compute the topological charge.
Since for every twist we have found a particular set of twist matrices---the abelian 
ones---, we might compute
the value of Q for this class of solutions. 
By taking  Eqs.~\ref{Omabelian},\ref{omegamunu},~\ref{Tmunu} and substituting it into 
Eq.~\ref{topch}, we obtain:
\be
\label{topchab}
Q= - \frac{\kpn}{N} + \frac{1}{4} Tr( {\bf Q}_{\mu \nu}  {\bf
\tilde{Q}}_{\mu \nu})  = - \frac{\kpn}{N} + Tr(\vec{\EPS}\vec{\BET})\ .
\ee
The second term  is an integer. For example, had we taken the particular solution
Eq.~\ref{partabetm}, we would have gotten $Q=\kpn \frac{N-1}{N}$.
Actually, with a
suitable choice of  $\EPS_i \equiv {\bf Q}_{0 i }$ and
$\BET_i \equiv \frac{1}{2}{\bf Q}_{j k  }\epsilon_{i j k }$,
one can make the second term of Eq.~\ref{topchab} take
any possible integer value $n$. A possible choice that does the job is
($i=1,2$):
\begin{eqnarray}
&\EPS_i =\left( \begin{array}{c c c c} -k_i & 0 & \ldots &0 \\ 0 & 0 & & \\
\vdots & \vdots & \ddots & \\ 0 &0 & & 0 \end{array} \right) \ \ 
&\BET_i =\left( \begin{array}{c c c c} 0  & 0 & \ldots &0 \\ 0 & -m_i & & \\
\vdots&\vdots & \ddots & \\ 0&0 & & 0 \end{array} \right) \ \  \\
\nonumber
&\EPS_3 =\left( \begin{array}{c c c c c} -k_3-n  & 0 & 0& \ldots &0 \\ 0
& 0 &0 & & \\ 0 & 0 & n & \ldots & \\
\vdots & &   &\ddots & \\ 0& & & & 0 \end{array} \right) \ \ 
&\BET_3 =\left( \begin{array}{c c c c c} 0  & 0 & 0 & \ldots &0 \\ 0 & -m_3-1
&0&  & \\ 0 & 0 & 1 &  \ldots & \\
\vdots &&  & \ddots & \\0  & & & & 0 \end{array} \right)\ .
\end{eqnarray}
The previous formula makes sense  for $N \ge 3$. For $\SUT$
and  no-twist ($\vec{k }=\vec{m}=0 \bmod 2$), there is an exception
to the rule: Only even values of the 
topological charge are attainable in this case.

In summary, we have shown that in the 
presence of twist, the topological  charge is given by: 
\be
\label{topchval}
Q= - \frac{\kpn}{N}  + n \ , \ \ n \in {\Z}\ .
\ee
The previous formula was conjectured by `t Hooft~\cite{twist1,twist2}, and
actually proven along similar lines to our presentation by Pierre van
Baal~\cite{vanbaal1}.  A consequence one can derive from Eq.~\ref{topchval}
is that for non-orthogonal twists
($\kpn \neq 0 \bmod N $) $Q$ is not an integer.  For orthogonal twists, one has 
twist-eating solutions, whose topological charge  vanishes. 


Now let us comment  about the appropriate mathematical
explanation of our results. The topological charge for an
$\SUN$ group, is just minus the second chern number, which is an integer. 
In view of this, the previous result is perplexing. However, as
mentioned previously, we have to look  at our fields and bundles as being
defined in
$\SUN/\ZN$. Hence, we should  rather compute the previous expressions in the
adjoint representation, which is a faithful representation of $\SUN/\ZN$.
To do so, we  have to replace in Eq.~\ref{topch} $\LAa$ by $\Lada$.
The result gets multiplied by $\frac{c_A d_A}{c_F d_F} = 2 N$, where
$c_{A,F}$ and $d_{A,F}$ are the quadratic Casimir and the dimension 
of the adjoint and fundamental representations. We get:
\be
Q^{(adjoint)} = - 2 \kpn + 2 N n \ .
\ee
which is an integer. The previous quantity is more appropriately referred
to as the first Pontryagin index $p_1$. Indeed, mathematicians have proven
that $p_1$ and twist provide a complete classification of the
bundles~\cite{sedlacek}.

\section{from $\UN$ to $\SUN / \ZN$}
In this section, we will look at   twist from   a different point of view, by
constructing a twisted $\SUN$ gauge field as a projection of a   $\UN$ gauge
field one\cite{lpr}. Our starting point, hence, is a $\UN$ gauge field on the torus:
$ {\bf A}'_{\mu}(x) = A{'}_{\mu}^{a}(x)  \ \LAa \ $. The difference 
with respect to the  $\SUN$ field case,
is the presence of  an additional
generator   $\boldsymbol{\lambda}_0=\frac{\mathbf{I}}{\sqrt{ 2 N}}$.
This field satisfies the boundary condition:
\be
\label{tbcUN}
{\bf A}'_{\mu}(x+\nh) = \BPL\Omega'_{\nu}(x) \BPR {\bf A}'_{\mu}(x)\ ,
\ee
where now the matrix $\Omega'_{\nu}(x)$ belongs to $\UN$.
The compatibility condition satisfied by these matrices is:
\be
\label{twisteqUN}
\xOrhonph =   \xOnurph\  \ .
\ee

It is now possible to decompose the gauge field into an $\SUN$ part $\Amu$ and
a $\UO$ part, by considering the corresponding components of ${\bf
A}'_{\mu}(x)$. The boundary condition Eq.~\ref{tbcUN} somewhat mixes the two
parts. However, we might factor the matrices $\Omega'_{\nu}(x)$ into an $\SUN$
 and a $\UO$ part as follows:
\be 
\Omega'_{\nu}(x) = \exp\{\imath\, \omega_{\nu}(x)\}\ \Omega_{\nu}(x)\ \ ,
\ee
where $\Omega_{\nu}(x) \in \SUN$
The decomposition is unique modulo multiplication of $\Omega_{\nu}(x)$
by an element of the center $z_{\nu}$ accompanied by the corresponding
shift in $\omega_{\nu}(x)$. However, by continuity  $z_{\nu}$ is space independent, 
and does not influence the following. Now, for the 2 different parts 
we have the following boundary conditions:
\begin{eqnarray}
&{\bf A}_{\mu}(x+\nh) = \Onu {\bf A}_{\mu}(x)\ \\
& A{'}_{\mu}^{0}(x+\nh) =  A{'}_{\mu}^{0}(x) + \sqrt{2 N}\
\partial_{\mu}\, \omega_{\nu}(x)\ \ . 
\end{eqnarray}
The compatibility conditions read:
\begin{eqnarray}
\xOrhonh = \exp\{2 \pi \imath \frac{n_{\rho \nu}}{N}\} \  \xOnurh\ \\
\label{primer}
\Delta_{\rho}\omega_{\nu}(x) - \Delta_{\nu}\omega_{\rho}(x) = \frac{2
\pi}{N} n_{\rho \nu} \bmod 2 \pi   \ .
\end{eqnarray}
 Henceforth, in general, the resulting $\SUN$ field has a  non-trivial twist
 tensor $n_{\rho \nu}$. We can compute the field-strength tensor and
 decompose it into the $ \SUN$ and $\UO$ part:
 \be
{\bf F}'_{\mu \nu}(x) =  {\bf F}_{\mu \nu}(x) +   F'^{0}_{\mu \nu}(x)\ 
\boldsymbol{\lambda}_0
 \ee
where  $F^{0}_{\mu \nu}(x)  =\partial_{\mu} A{'}_{\nu}^{0}(x)  - \partial_{\nu}
A{'}_{\mu}^{0}(x)$.

 Now, we could use these results to relate the topological properties of
 the $\UN$ field to those of $\SUN$ gauge fields with twist. It is well-known 
 from the theory of characteristic classes\cite{egh} that the topological
properties of these  fields (actually of the bundles or transition matrices) 
are characterised by the Chern classes. In our case, we have the following 
topological invariants:
\begin{eqnarray}
c_{\mu \nu} = \frac{1}{2 \pi} \int dx_{\mu}\wedge dx_{\nu}\ Tr({\bf F}'_{\mu
\nu}) \\
k= \frac{1}{16 \pi^2}\int_{T_4} Tr(  \Fpmunu\, \Ftpmunu)\, d^4x\ \ .
\end{eqnarray}
The first set of invariants $c_{\mu \nu}$ is the integral of the first Chern
class over the non-contractible 2-dimensional surfaces of the torus. These
numbers are integers and hence do not depend on the actual surface chosen
but only on the directions $\mu$ and $\nu$. The other quantity $k$ is again 
an integer and is
obtained from the second Chern class. Now, we can evaluate these invariants
in terms of our decomposed fields:
\begin{eqnarray}
c_{\mu \nu} = \frac{N}{2 \pi} (\Delta_{\nu}\omega_{\mu}(x) -
\Delta_{\mu}\omega_{\nu}(x)) \\
\label{second}
k= Q + \frac{1}{4 N }\ c_{\mu \nu}\, \tilde{c}_{\mu \nu}\ \  ,
\end{eqnarray}
where we have used the boundary conditions and compatibility equations.
Notice, that using Eq.~\ref{primer} one arrives at:
\be
n_{\mu \nu} = c_{\mu \nu} \bmod N \ \ ,
\ee
which relates the twist tensor  with the  first  Chern class of the original
 $\UN$ field. Given this,  Eq.~\ref{second} reproduces the expression of
the topological charge $Q$ of an $\SUN$ field with twist that we  found
before (Eq.~\ref{topchval}).

\section{Gauge invariant Quantities}
The most important quantities of the theory are the Wilson loops. They 
are given by:
\be 
W(\gamma) = Tr(U(\gamma:P\rightarrow P))\ ,
\ee
where $U(\gamma:P\rightarrow P)$ is the parallel transporter matrix
along the closed loop  $\gamma$. Closed paths on the torus can be classified
into the different homotopy classes, which form the group $\pi_1(\TF)=\ZN^4$.
An element $w=(w_{\mu})$ of this group contains all paths which
wind $w_{\mu}$ times around the $\mu$-th direction of the torus.
The corresponding Wilson loops are called Polyakov loops or Polyakov lines.
The simplest of these, are straight line Polyakov loops along the  $\mu$-th
direction:
\be 
\label{Polline}
P_{\mu}(x)\equiv Tr(U_{\mu}(x)) = Tr ( T\exp{ \{-  \imath \int_{0}^{l_{\mu}} \Amu\, dx_{\mu} \}}\: 
\Omega_{\mu}(x,x_{\mu}=0) )\ .
\ee
The ocurrence of the twist matrix $\Omega_{\mu}$ in the previous formula, is essential 
to guarantee gauge invariance.
In general, one should insert a twist matrix every time that the loop $\gamma$
traverses the boundary of the patch. 

 Polyakov lines are non-invariant under the $\ZN^4$ discrete group of
 transformations (Eq.~\ref{ZNsym}). If  $k\equiv(k_{\mu})$ is an element of this group,
 the Polyakov loops transform $W(\gamma) \rightarrow \exp{ \{2 \pi \imath\: k
\cdot w(\gamma) / N \}}\, W(\gamma)$. It is also important to notice that 
Polyakov lines are non-periodic. For example, from its definition
Eq.~\ref{Polline} one can easily deduce  that:
\be
P_{\mu}(x+\nh)= z_{\mu \nu} P_{\mu}(x)\ .
\ee

\section{Classical Solutions}
This section will try to summarize what is known about solutions 
of the euclidean classical equations of motion of \YM\ fields 
on the torus. Classical solutions are extrema of the \YM\ (euclidean)
action functional:
\be
S= \frac{1}{2} \int_{\TF} d^4x\, Tr(\Fmunu\, \Fmunu) \ .
\ee
They satisfy $D_{\mu} \Fmunu = 0$. Some of the classical solutions are those 
which minimize the action within each twist and topological charge sector, 
and are referred to as  {\em minimum action solutions}.
Using the Schwartz inequality one can
derive the following bound  for $S$:
\be 
S \geq  8 \pi^2 |Q| \ .
\ee
The bound is saturated by  self-dual and anti-self-dual solutions ($\Fmunu =\,
\pm\, \Ftmunu$), provided they exist.
For non-orthogonal twists, the topological charge is never an integer,
and  henceforth the action can never be zero. 

It is clear that given one classical solution, one can obtain others for the
same twist, topological charge and action, by making a gauge transformation.
They are said to be gauge-equivalent. Our  purpose would be to find all
gauge-inequivalent classical solutions. In the next sections we will review  
the different cases which have been studied in the literature.   

\subsection{Zero-action solutions}
As mentioned previously, zero action solutions ($S=0$)  can only occur for
orthogonal twists, which  is the class of twists for which $\vec{k}\cdot\vec{m}
= \kpn = 0 \bmod N$. 
If we take a zero vector potential $\Amu = 0$, then this is only compatible
with the boundary conditions Eq.~\ref{tbc} provided the twist matrices are
constant. Indeed, this is the class of twist matrices $\Gmu$ which was studied
previously under the name of twist eaters. It is now clear what is the
origin of the name. 

Having found zero-action  solutions for all orthogonal twists, we turn now
to the problem of determining how many gauge-inequivalent sets are there.
Within the set of  $\Amu = 0$ configurations, one is still free to make
global gauge transformations, which will not change the value of the vector 
potential. However, these operations transform  the twist matrices by a
similarity transformation. Henceforth, the set of inequivalent twist-eaters 
labels the gauge-inequivalent configurations of this type.  

Alternatively, one might take  fixed twist matrices $\Omegamu = \Gmu$, and
consider different  values of the vector potential. Since $S=0 \Rightarrow
\Fmunu = 0$,  there must exist $\Omega(x)$ such that: 
\be 
\Amu = \imath\, \Omega(x) \partial_{\mu} \Omega^+(x)\ .
\ee
However, the vector potentials must satisfy the boundary condition 
Eq.~\ref{tbc}, which in our case demands:
\be
\Omega(x+\nh) \partial_{\mu} \Omega^+(x+\nh) = \Gnu\,  \Omega(x) \partial_{\mu}
\Omega^+(x)\, \Gnu^+ \ . 
\ee
This can be shown to imply:
\begin{eqnarray}
\label{consttild}
&\partial_{\mu} \tilde{\Gamma}_{\nu}(x)=0\ , \\
 \mbox{where}\  \ &\tilde{\Gamma}_{\nu} = \Omega^+(x+\nh) \Gnu \Omega(x)\ . 
\end{eqnarray}
Now the question  is whether this is simply a gauge transformation of the
previous  ( $\Amu = 0$) solution. If we perform  a gauge transformation with
$\Omega^+(x)$,
we  indeed get $\Amu \rightarrow  0\,;\ \Gmu \rightarrow \tilde{\Gamma}_{\mu}$.
The new twist  matrices $\tilde{\Gamma}_{\mu}$ are constant as a consequence
of Eq.~\ref{consttild}, and we are back to the previous situation. In
summary, we have proven that:  
\begin{itemize}
\item The class of gauge inequivalent zero-action solutions coincides with the class
of inequivalent  twist-eaters. 
\end{itemize}
Let us now characterize the elements of this class in terms of gauge
invariant quantities. If we evaluate all Wilson loops for the configuration
with zero vector potential and twist matrices equal to $\Gmu$, we get 
\be
W(\gamma)= Tr ( \Gamma^{\epsilon_{1}}_{\mu_{1}} \cdots
\Gamma^{\epsilon_{s}}_{\mu_{s}})= e^{\imath \alpha(\gamma)}\,
Tr(\Gamma(\omega(\gamma)))\ , 
\ee
where s is the number of times the path $\gamma$ crosses the edge of
the hypercube $\prod_{\mu}[0,l_{\mu}]$ and $\alpha(\gamma)$ a path-dependent phase.
Every time the curve $\gamma$  crosses the
edge of the hypercube along  the $\mu$th
direction one gets a  factor of $\Gmu^\epsilon$ in the previous expression,
with $\epsilon = \pm 1 $ depending on the sense of crossing.
The final trace is only non-zero when $\Gamma(\omega(\gamma))$ is an element
of the center $\cal C$ of the twist group.
Henceforth, the  Casimirs of the twist group $\cal G$ correspond to non-zero Polyakov lines.

We have  set the mathematical problem properly: the space of gauge
inequivalent zero-action solutions (or euclidean vacua),
which is  known as  {\em the euclidean vacuum
valley}, is the space of equivalence classes of N-dimensional
representations of the twist group (In the mathematical literature, 
zero-action solutions are referred to as {\em flat connections}). In what follows, we will study the structure of this space 
for the different twists. The study naturally splits into the reducible and
irreducible twist cases.

\subsubsection{Irreducible twists}
Under this name we include the values of $\nmunu$ and $N$ for which we have
an irreducible representation of the twist group. We recall that this occurs
when $p_1 \cdot p_2 = N$ for rank 2, or $p_1=N$ for rank 1.  In either case
we proved previously that the number of inequivalent solutions is discrete.
Indeed, we showed that the number of such solutions is $N^2$. There is
always a Polyakov line which takes a different value in any two of these vacua.
 As a particular example, consider the twist $\vec{m} = (1,1,1); \vec{k} =
(0,0,0)$.  This has rank 1 and $p_1=N$.  The two independent Polyakov lines
are $P_0=Tr(U_0)/N$ and $P_{(1,1,1)}=Tr(U_1  U_2 U_3 )/N$.
Each one can take N different values ( the $N^{th}$ roots of unity).

Another example of rank 2 is  $\vec{m} = (n,0,0); \vec{k} =(\frac{N}{n},0,0)$
(n divides N).  In this case the independent Polyakov lines are:
$Tr(U_2^{N/n})/N$, $Tr(U_3^{N/n})/N$, $Tr(U_0^n)/N$ and
$Tr(U_1^n)/N$. The first 2 can take n values and the last two $N/n$ 
values.

\subsubsection{Reducible twists}
This case is much more complicated than the previous one. We have $N= s N_0
 $, where s is the number of irreducible components. One can bring the
matrices to block diagonal form. In each $N_0 \times N_0$ box one has an
irreducible representation of the twist group. Since unitary irreducible 
representations are unique modulo multiplication by a phase, we might 
write all the matrices in the form:
\be
\Gmu =  S_{\mu} \otimes \Gmu^{(N_0)}\ ,
\ee
where $\Gmu^{(N_0)}$ are the matrices of one  irreducible representation,
and $S_{\mu}$ are diagonal SU(s) matrices.

The case that we will study first is the purely periodic one ($\nmunu = 0$).
This is a particular case of the general reducible case having   $s = N$ and
$\Gmu= S_{\mu}$. It might seem that the
eigenvalues (elements) of $S_{\mu}$ label the different non-gauge equivalent solutions.
In that case, the vacuum valley would be isomorphic with $S_1^{dN}$, where
d=4 is the space-time dimension.  However,
this is not so, because there are different points of this manifold which
are gauge equivalent.  This occurs because there are similarity
transformations which bring the $S_{\mu}$ back to diagonal form, namely those
which exchange the order of the eigenvalues simultaneously in all d matrices.
An additional complication comes from the requirement that the matrices have
determinant equal to 1. This conditions fixes the value of the last
eigenvalue in terms of the previous ones. Hence, we have the manifold $S_1^{d
(N-1)}$. Now we may introduce the group of transformations $\cal T$ in this 
manifold  corresponding to an exchange of the N eigenvalues. This group is 
isomorphic to the permutation group of N elements. Hence our vacuum valley is:
\be
 {\cal V} \equiv S_1^{d(N-1)}/{\cal T} \ .
\ee
The disgusting feature of this space, is that it is an orbifold and not a
manifold, corresponding to the fact that there are fixed points of the group
of transformations $\cal T$. They occur   when there are degenerate 
eigenvalues  for all $S_{\mu}$ at the same time. 

The presence of an infinite number of gauge-inequivalent vacua for periodic
gauge fields on the torus was pointed out first in Ref.~\cite{g-ajk-a}. It was realized that
there are special points in this space corresponding to the singular points.
The name {\em toron} was used to refer to  a gauge-equivalence class of vacua.
In that reference, instead of setting $\Amu = 0$ and labelling the torons by
the twist matrices, the latter were fixed  to unity, and each toron was
labelled by the value of $\Amu$. The transformations which leave the unit twist
matrices invariant are the periodic gauge transformations. It is possible to
choose one representative of each class of gauge-equivalent vector
potentials, within the set of constant diagonal vector potentials. This is an 
alternative and equivalent description of the vacuum valley. 

If we go to the generic irreducible case, our analysis and description
remains valid after the substition $N \rightarrow s$. Henceforth, the
euclidean vacuum   valley ${\cal V}$ is given by:
\be
 {\cal V} \equiv S_1^{d(s-1)}/{\cal T}_s \ ,
\ee
where  ${\cal T}_s$ is isomorphic to the permutation group of s elements.

\subsection{Abelian solutions}
The first class of solutions which were  known for orthogonal and 
non-orthogonal twists  are abelian.
Consider first the abelian twist matrices given in
Eqs.~\ref{Omabelian},~\ref{omegamunu},~\ref{Tmunu}.
If $\Fmunu$ and $\Amu$ are chosen to conmute  with these matrices, the vector potential 
should satisfy the following boundary condition:
\be
\Delta_{\nu}\Amu = \partial_{\mu} \omega_{\nu} = \frac{\pi}{N} \frac{T_{\nu
\mu}}{l_{\mu}}\ . 
\ee
A solution of this equation is given by:
\begin{eqnarray}
\label{Acfs}
\Amu &=& -\frac{\omega_{\mu}}{l_{\mu}} = 
- \frac{\pi}{N}  \frac{x_{\nu}}{l_{\nu} l_{\mu}} {\bf T}_{\mu
\nu}\ \\
\Fmunu &=& \frac{2 \pi}{N} \frac{T_{\mu \nu}}{ l_{\mu} l_{\nu}}
\label{Fcfs}
\end{eqnarray}
This is indeed a classical solution ($D_{\mu} \Fmunu =
\partial_{\mu} \Fmunu = 0$). We can compute for this configuration 
the value of the action and the topological charge:
\begin{eqnarray}
\label{actionab}
S&=& \frac{2 \pi^2}{N^2}\, l_0 l_1 l_2 l_3\, Tr\left( \frac{T_{\mu \nu}}{l_{\mu}
l_{\nu}}\right)^2 = \frac{ 4 \pi^2}{N^2} \sum_{i=1}^{3}\, [q_i\, Tr({\cal E}_i^2) +
\frac{1}{q_i} Tr({\cal B}_i^2) ] \\
Q &=& \frac{1}{4 N^2}\, Tr(T_{\mu \nu} \tilde{T}_{\mu \nu}) = \frac{1}{N^2}
Tr(\vec{\cal E} \vec{\cal B})\ ,
\label{qab}
\end{eqnarray}
where $T_{0 i }={\cal E}_i$, $T_{i j  }=\epsilon_{i j k} {\cal B}_k$ and
\be
q_i = \frac{l_0 l_1 l_2 l_3}{l_0^2 l_i^2}\ .
\ee

As  expected, the value of the topological charge is the one which follows
from the twist matrices. Therefore, there are  solutions for all values of
the twist  and  the topological charge (excepting the odd integers for
SU(2) with twist $\vec{k}= \vec{m} = 0 \bmod 2$). The corresponding value of the
action follows from Eq.~\ref{actionab}. Here we will look in bigger
detail to the self-dual (or anti-self-dual) solutions, which when existing 
have the  minimum possible action within each twist and topological charge sector.
 The self-duality  condition
implies $q_i {\cal E}_i = {\cal B}_i$. As a consequence, the $q_i$ have to be 
rational numbers:
\be
\label{lenratios}
q_i = \frac{p_i}{p'_i}\, ,\ \ \mbox{ with } p_i,p'_i \in \Z^+ \ ,  
\ee
and  $p_i$ and $p'_i$ coprime. Henceforth, we might express ${\cal E}_i =
p'_i\, 
\mathbf{K}_i$ and  ${\cal B}_i =p_i\,  \mathbf{K}_i$, where $\mathbf{K}_i$ 
are traceless, integer, diagonal matrices. One gets:
\be
Q= \sum_{i=1}^{3}\, \frac{1}{N^2}\,  p_i p'_i\ Tr(\mathbf{K}_i^2)\ .
\ee
We still have to put in the condition Eq.~\ref{Tmunu}. Since there exist
integers $a_i$ and $b_i$ such that  $a_i\, p_i+b_i\, p'_i=1$ ($i$ fixed), we may
solve for  $\mathbf{K}_i$:
\be
\label{Kabe}
\mathbf{K}_i = s_i + N\, \mathbf{S}_i\ ,
\ee
where $\mathbf{S}_i$ are integer, diagonal matrices satisfying
$Tr(\mathbf{S}_i) = -s_i$, and $0 \le s_i < N$ is such that $k_i=p'_i
\, s_i \bmod N$ and $m_i=p_i\, s_i \bmod N$. The topological charge becomes: 
\be
\label{Qsdabe}
Q= \sum_{i=1}^{3}\, p_i\, p'_i\, \left( -\frac{s_i^2}{N} +  Tr(\mathbf{S}_i^2) \right) \ge 0 \ .
\ee
The first term is up to an integer equal to $-\frac{\kpn}{N}$.
For fixed size and twist the minimum value of the topological charge
is $Q=\sum_i p_i\, p'_i\, s_i (N-s_i) /N$.  If we are given the twist ($0
\le  k_i, m_i < N $), it is always possible to find a given size of the torus 
for which there is an abelian self-dual solution.  

We might summarize our results in the following statements:
\begin{itemize}
\item There are abelian solutions for all sizes, twists and topological
charges, except  for $\SUT$, no-twist and odd $Q$.
\item Self-dual ( $Q \ne 0$) solutions occur for all twists and for some
torus sizes, such that the  ratios of the lengths in  each direction
are square roots of rational numbers: $l_i/l_0 = \sqrt{q_1 q_2 q_3 /q_i}$.
The minimum value of the topological charge for this configuration
is $Q=\sum_i\,  p_i\, p'_i\, s_i (N-s_i) /N$, where  $p_i$ and $p'_i$ are
defined in Eq.~\ref{lenratios} and $s_i=a_i\, m_i + b_i\, k_i$ (with
$a_i\, p_i+b_i\, p'_i=1$). 
\item The only case in which these abelian solutions are minimum action
solutions within each twist sector, is for $\SUT$ and   $\vec{k}=\vec{m}=(1,0,0)$ ($p_i=p_i'=1$). 
By a misfortunate coincidence, this configuration was called {\em toron}
by the authors of Ref.~\cite{ec+cg} ---the same name used before for a 
different object---~.
\end{itemize}

The anti-self-dual solutions have negative values of the topological charge
and can be obtained from the previous ones by time reversal or parity. 

\subsection{Non-abelian constant  field-strength solutions}
`t Hooft~\cite{stsds} discovered  that a wider class of 
constant conmuting $\Fmunu$
solutions are compatible with the twisted boundary conditions.
To find them, let us work the other way round and start with the expression of the
vector potential and of the field strength tensor Eqs.~\ref{Acfs}-\ref{Fcfs}.
We will try  to find out what are the possible twist matrices $\Omegamu$ in
this case. They must conmute with $T_{\mu \nu}$. From the boundary conditions
for $\Amu$ (\ref{tbc}) one gets:
\be
\imath\, \Omeganu \partial_{\mu} \Omegatnu = \Delta_{\nu}\Amu =  \frac{\pi}{N} \frac{T_{\nu \mu}}{l_{\mu}} 
\ .
\ee
The general solution of the previous equation is given by:
\be
\Omegamu = \exp \{\imath\,  \frac{\pi  }{N }\,
\frac{x_{\nu}}{l_{\nu}}\, {\bf T}_{\mu \nu}
\}\, \Gmu\ , 
\ee
where $\Gmu$ are constant $\SUN$ matrices which conmute with ${\bf T}_{\mu
\nu}$. These matrices satisfy: 
\be
\Gmu \Gnu \Gmu^+ \Gnu^+ = \exp{\{ - 2 \pi \imath\, {\bf Q}_{\mu \nu}'    \}
}\ .
\ee
The form of  ${\bf Q}_{\mu \nu}' $ follows from our previous study of twist
eaters. If there are $N_{1}$ coinciding eigenvalues in ${\bf T}_{\mu \nu}$,
we might set $\Gmu$ within this subspace to be one of the twist eating
solutions. Hence, in this subspace, one has that the eigenvalues of
${\bf Q}_{\mu \nu}'$ are given by $q'^{(1)}_{\mu \nu}/ N_1$, with $q'^{(1)}_{\mu
\nu}$ an orthogonal twist tensor in $SU(N_1)$. Now imposing the twist 
condition Eq.~\ref{twisteq} we get:
\be
 {\bf T}_{\mu \nu} = \nmunu\, {\bf I} + N\, {\bf Q}_{\mu \nu}'\ .
\ee
The difference with respect to the   purely abelian case lies in the
expression of ${\bf T}_{\mu \nu}$ (versus Eq.~\ref{Tmunu}), since ${\bf Q}_{\mu \nu}'$ is not
an integer matrix. With this difference, expressions~\ref{actionab}-\ref{qab}
are valid in this case as well. If we focus on self-dual solutions,
the main difference with the previous case is the form of $\mathbf{K}_i$, 
in which Eq.~\ref{Kabe} is replaced by:
\be
\mathbf{K}_i = s_i + N\, \frac{s_i^{(\alpha)}}{N_{\alpha}}\,
\boldsymbol{\delta}_{\alpha}\ .
\ee
What we have done, is to separate the N eigenvalues into sets of $N_{\alpha}$ coinciding
ones, so that $\sum_{\alpha} N_{\alpha} =N$.
The matrix $\boldsymbol{\delta}_{\alpha}$ is the projector onto the subspace
${\alpha}$, and $s_i^{(\alpha)}$ are integers such that $\sum_{\alpha}\, s_i^{(\alpha)}= -s_i$.
The  formula for the
topological charge   is now:
\be
Q= \sum_{i=1}^{3}\, p_i\, p'_i\, \left( -\frac{s_i^2}{N} +
 \sum_{\alpha} \frac{(s_i^{\alpha})^2}{N_{\alpha}} \right) \ge 0 \ .
\ee
The abelian  formula  Eq.~\ref{Qsdabe} is a particular
case of this one for $N_{\alpha}=1 \ \forall \alpha= 1 \ldots N$.
The condition that the twist in the $\alpha$ subspace be orthogonal,
amounts simply to the fact that the contribution of this space to the
previous formula be an integer ($\sum_i\, p_i\, p'_i\,
\frac{(s_i^{(\alpha)})^2}{N_{\alpha}} \in {\bf Z}$). 

The particular case studied explicitly by `t Hooft is that of 2
subspaces $N=N_1 +N_2$. We can work out explicitly the topological charge
for this case, and we get:
\be
\label{Qsplit2}
Q= \sum_{i=1}^{3}\,  p_i\, p'_i\, \frac{(s_i^1 \, N -s_i\, N_1)^2}{N_1\, N_2\, N}
\ee
In this case,  as shown by `t Hooft~\cite{stsds}, one can get  solutions which have the minimum action within
each twist sector.
By taking $p_i = N_1$, $p'_i=N_2$, $s_{1,2}=s^1_{1,2}=0$, and $s_3^1 \, N
-s_3\, N_1 =1$, one
gets $Q=1/N$. Other choices are easily extracted from Eq.~\ref{Qsplit2}.

\subsection{General non-abelian minimum action solutions}
Concerning non-abelian solutions in arbitrary  sizes and topological charges
not much is known analytically. Mathematicians have given existence proofs 
of some of these configurations~\cite{taubes}. However, a good deal is known about the
form and properties of these solutions by numerical methods. We will in what 
follows  give a brief description of these results without entering too much
into the technical details of the method employed. For that we refer the
reader  to the original papers \cite{gpg-as,gpg-a,new}.

\subsubsection{$\SUT$}
The numerical  work in this case has concentrated in studying the
minimum action solutions for the most spatially-symmetric choice of twist
$\vec{m}=\vec{k}=(1,1,1)$ and  size ($l_1=l_2=l_3=l_s$). In this case,
$\kpn=\frac{1}{2}$ and the action of all  configurations must be larger
than $4 \pi^2$. The numerical method used by the authors of Ref.~\cite{gpg-as}
is based on the lattice  gauge theory formulation of Wilson~\cite{Wilson}. 
This is a gauge-invariant discretization of the  Yang-Mills theory. As
mentioned previously, putting the system on a box with periodic boundary 
conditions amounts to the discretization of a toroidal base space.
Furthermore, it is possible to give in
this case a lattice counterpart of the \tbc~\cite{gjk-a,tek2}.
Although the lattice formulation destroys some of the topological features
of the continuum theory, this is not the case for twist.
Then, the point is to find by numerical methods the configuration which
minimizes a discretized version of the  continuum action-functional.
The authors have used different discretized versions, but most of their 
results are based on  Wilson's discretized form. By employing a 
so-called classically improved lattice action, one makes the discretization 
errors  smaller, but unfortunately the numerical techniques become more 
time consuming. In order  to obtain the minimum lattice action
configuration, the authors of Ref.~\cite{gpg-as,gpg-a}  used a local
minimization method known as {\em cooling}~\cite{cooling1,cooling2}. 
Finally, they extracted discretized estimates of the relevant quantities
and studied their convergence towards the continuum limit as the  lattice
becomes finer. 

The lattices are hypercubic and with  sizes which  were always  symmetric 
in space ($N_s^3 \times N_t$),
where $N_s$ and $N_t$ are the number of points in the spatial and temporal
directions. If one fixes the
continuum unit of length by setting  $l_s=1$, then the lattice spacing is
given by $a=\frac{1}{N_s}$.

\begin{figure}
\caption{{
We plot the value of the lattice action $S$ divided by $4 \pi^2$ as a
function of the square lattice spacing $a^2$,
for the SU(2) configuration which minimizes the action with twist
$\vec{m}=\vec{k}=(1,1,1)$.
The solid line is a linear fit to the data.
} }
 \hbox{ \hskip2.5truecm \epsfxsize=3.5truein \hbox{\epsffile{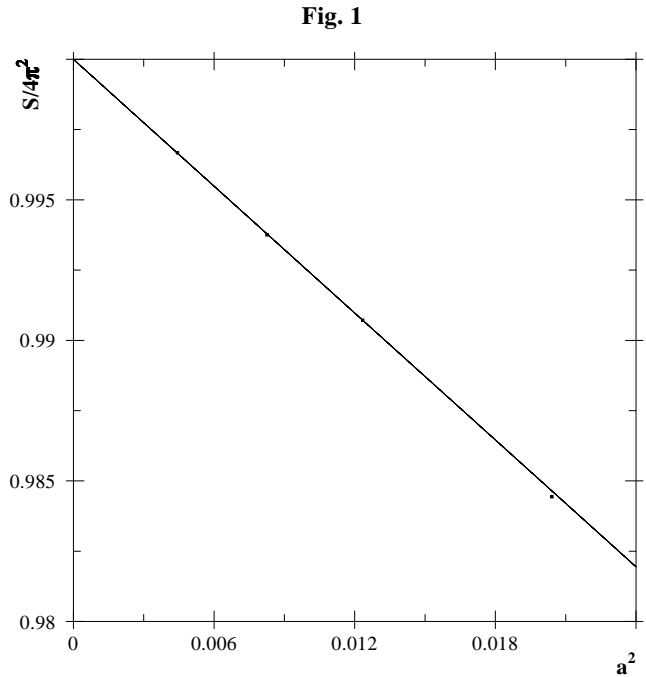}
} }

     \label{fig1}
     \end{figure}
     
The simplest  quantities to study are global
ones, such  as the total action and topological charge.    Fig.~1 shows
the value of the Wilson action for different values of $a$ and  $N_t/N_s \rightarrow
l_t/l_s \approx 2$. The data approaches  $4 \pi^2$ linearly in $1/a^2$. The
slope depends on the lattice action used, but the extrapolation can give
very precise results. From the data of Fig. 1 one gets an extrapolated
continuum action of $4 \pi^2$ with a precision of 1 part in $10^5$, which is
consistent with the size of the errors from other sources.
 At the same time, the topological charge tends to $Q=\frac{1}{2}$ 
 with equivalent precision. This is a strong 
indication that one is indeed approaching a self-dual classical
configuration. This is confirmed by studying 
self-duality directly (See Refs.~\cite{gpg-as,gpg-a,gpthesis} for details).
The other quantities studied also show 
that the lattice minimum configuration approaches a continuum one. The 
accuracy of the approximation depends on the sizes, the quantity studied,
the choice of discretized action and the choice of discretized estimate  of the 
different continuum quantities involved. In any case, for the values studied,
the numerical differences were at most of a few percent. In what follows, we 
will omit all reference to the method and concentrate on the results, taken as 
valid for the continuum minimum action configurations. 

The next quantity to study is the {\em energy profile} of the classical
solution, defined as:
\be
{\cal E}(t)= \int d^3x\, Tr(\vec{\bf E}^2(t,\vec{x}) +\vec{\bf B}^2(t,\vec{x}))\ .
\ee
For  $l_t/l_s > 1 $ this function has a maximum  at a given
time value, which will be called the time-center of the configuration, and
chosen to correspond to $t=0$. The 
main features shown by the energy-profile are the following:
\begin{itemize}
\item The curve approaches a well defined one when $l_t/l_s \rightarrow
\infty$.  
\item The curve is symmetrical about the time-center ($t=0$).
\item For large values of $|t|$, ${\cal E}(t)$ goes to $0$ exponentially.
\item The curves for finite $l_t/l_s > 1$  differ from the infinite
$l_t/l_s$ one mostly at the tails. Actually, the shape of the energy profile
is well described by a periodization of the $l_t/l_s \rightarrow \infty$ curve.
\item The same curve is obtained within errors if one takes twice the
electric part (or magnetic part) of ${\cal E}(t)$, in agreement with
self-duality.
\item To show the quantitive aspect of the function, we show in Fig.~2, 
the shape of a function which fits the numerical data for the  
$l_t/l_s \rightarrow \infty$ curve, with a
precision of the width of the line. The function has only 4 parameters
and its form is \[ F(t)=\frac{1}{0.0113 + 0.00323\, cosh(8.515 t)\,  +
0.0459 t^2} \ . \]
\end{itemize}
Notice that the coefficient  of the exponential decay is roughly equal to $2 \pi
\sqrt{2} \approx 8.88 $, which is the minimum curvature of the classical
potential.

\begin{figure}
\caption{{
We show the energy profile of the SU(2) minimal configuration with
$\vec{m}=\vec{k}=(1,1,1)$.
} }
 \hbox{  \hskip1.truecm  \epsfxsize=4.5truein \hbox{\epsffile{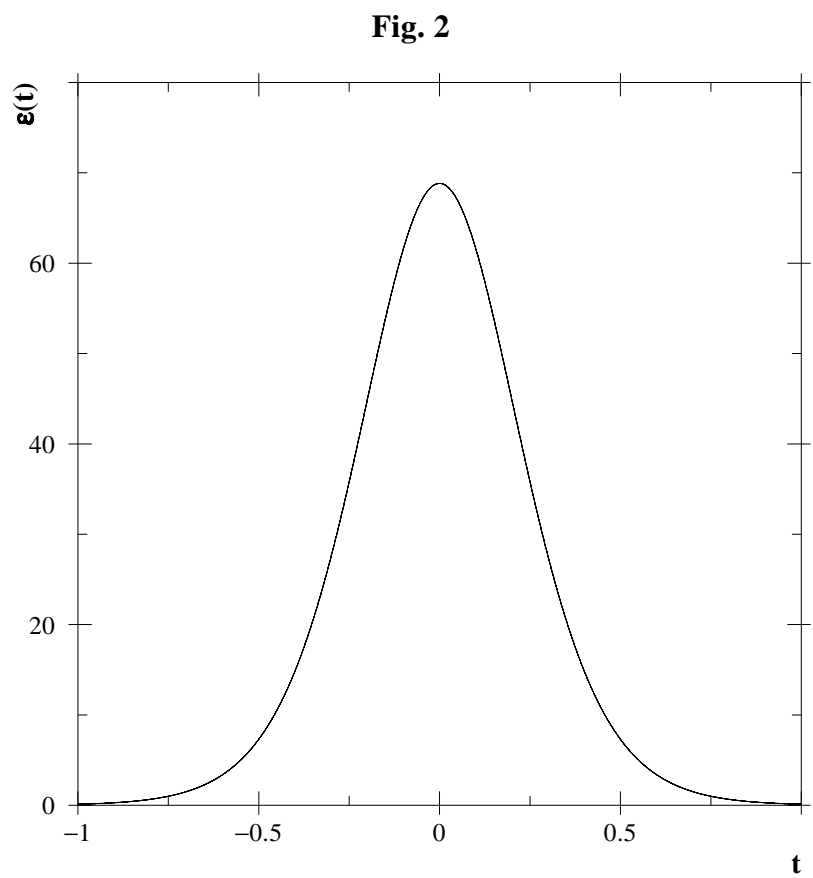}
} }
          \label{fig2}
	       \end{figure}

A qualitative conclusion is  that the $l_t/l_s \rightarrow \infty$ solution
can be considered an {\em instanton-like} solution, which  tunnels between
2 classical zero-energy states. The exponential decay at $\infty$ tells us
that the  system is not scale-invariant, being broken by the spatial size
of the torus. We will refer to this object as {\em $Q = \frac{1}{2}$
instanton}.  

In Refs.~\cite{gpg-as,gpg-a,gpthesis} the authors investigate in some detail the form
of  several physical   quantities for this configuration. The results are
obtained for a 
ratio $N_t/N_s = l_t/l_s \approx 2$ or higher, which except at the tails is already
very close to  $ \infty$. The field tensor $\Fmunu$,
the straightline Polyakov loops and the vector potential itself are studied.
The latter is obtained in the $A_0=0 $ ; $A_i(t=-\infty)=0$ gauge. The main
results obtained are:
\begin{itemize}
\item The configuration action density has a maximum at a given point in space and time
(center of the $Q = \frac{1}{2}$ instanton).
The size of the object (Full width at half maximum) is a large fraction of the
torus spatial size. 
\item The configuration is invariant under spatial rotations belonging to
the cubic group. This is the group which is left invariant by the boundary
conditions.
\item The solution is highly non-abelian. At the vicinity of the $Q =
\frac{1}{2}$ instanton center, the solution behaves similarly to the
ordinary BPST instanton. Hence, $E_1$, $E_2$ and $E_3$ are orthogonal and
of equal modulus in gauge space. In this sense it looks completely opposite 
to the constant field strength solutions (abelian and non-abelian),  for 
which these quantities are parallel in  Lie Algebra space.
\item Both the electric field(= magnetic field) and  vector potential
components can be nicely fitted with a few Fourier coefficients.
\item  The  space of solutions of this type ($Q = \frac{1}{2}$ for this twist)
depends on 4 parameters, which can be taken as the coordinates of the
center of the configuration. This, as will be commented later, is what the
index theorem predicts for this case. In addition,  there is a discrete degeneracy,
about which the index theorem can say nothing. Indeed, there are actually 8
gauge inequivalent families of instantons. They differ, for example, by the
sign of the straight Polyakov loops  passing through the $Q = \frac{1}{2}$ instanton center.
\end{itemize}

\subsubsection{$\SUN$}

In Ref.~\cite{new,gpg-am} a similar study is performed for other  SU(N) groups.
Again, the study
focuses in a spatially symmetric situation: $l_1=l_2=l_3=l_s$ and
$\vec{m}=(1,1,1)$. Different groups are studied from N=3 to N=25.
The reason for studying several groups will be more clear in the
next section. Actually, there are two cases which are particularly
interesting: $SU(3)$, because it is the physically relevant case for QCD,
and $\SUINF$, because there is a hope that the solution simplifies in that
limit and hence  a higher chance of achieving an analytical solution, which
is one of the goals  of the numerical study.
The temporal twist is chosen in several ways, but such that $|\kpn|=1$.
Let  us concentrate in this paragraph on  the results for $N$ equal or close
to $3$.
Essentially, the conclusions are similar to the SU(2) case, except that now
the cubic symmetry is replaced by the cyclic group $ x \rightarrow
y \rightarrow z \rightarrow x$. Another change is that at the $Q=\frac{1}{N}$
instanton center, the three spatial components of the electric field are
no longer orthogonal. In this sense, as $N$ grows the configuration looks
more abelian at the center. The shape of the energy profile does
not depend on the temporal twist (with $|\kpn|=1$) for $l_t/l_s$ large enough.
The  form of the
energy profile does not change too much from one group to another, provided
we scale the time variable by $1/N$ and profile itself  by $N^2$. The
resulting curves $N^2\,{\cal E}_N(t/N)$ are shown in  Fig. 3 for various groups.

\begin{figure}
\caption{{
Comparison of the energy profiles for different SU(N) groups.
The profiles are scaled as shown in the figure $N^2\,{\cal E}_N(t/N)$.
} }
\vbox{ \hbox{ \hskip1.truecm \epsfxsize=4.5truein \hbox{\epsffile{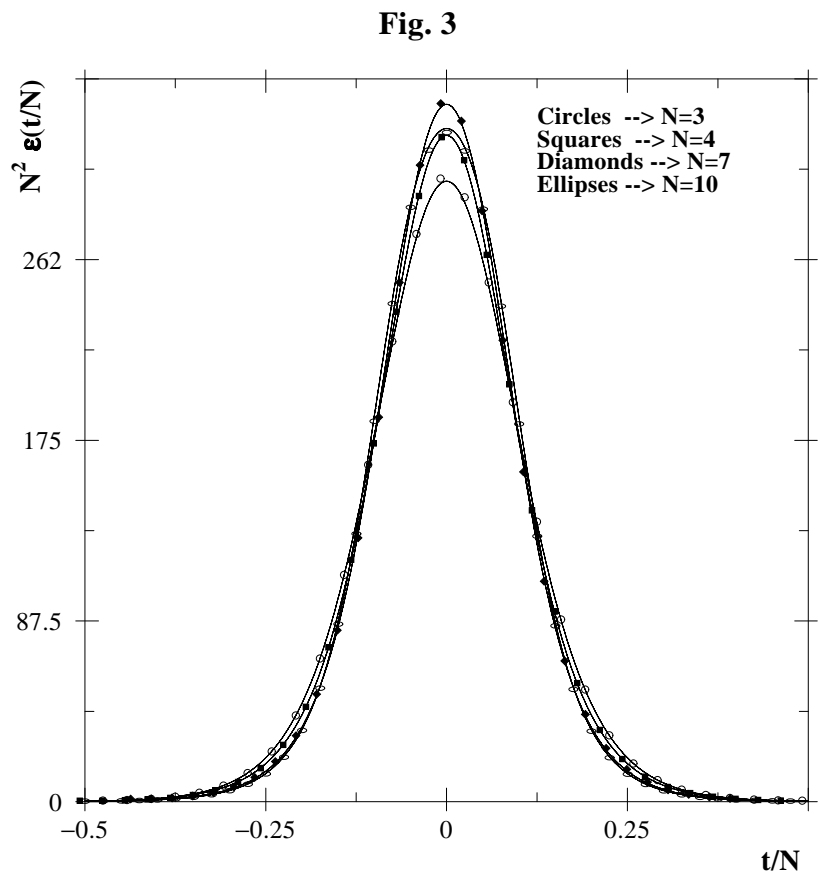}
} }
     }
               \label{fig3}
	                      \end{figure}
			      
\subsubsection{$\SUINF$}

The large $N$  limit is particularly interesting, since as `t
Hooft~\cite{stsds} noticed, the configurations which have an action going
like $1/N$ (unlike instantons),  are not suppressed as $N \rightarrow \infty$
in the path integral. Furthermore, as explained previously  there is a bigger
hope  to be able to obtain an analytical solution in this case.
Unfortunately, the approach towards  $N \rightarrow \infty$ in some
quantities is fairly slow. Exploring larger groups  numerically is limited by
computer memory and time. This slow tendency shows up, for example, in the
behaviour of the energy profile. For $N \approx 3$, as 
mentioned in the previous paragraph, its behaviour suggests that there could be a
well-defined limit of the function $N^2\,{\cal E}_N(t/N)$ as $N \rightarrow
\infty$. However, although numerically quite similar, there is no clear  way
to extrapolate to $\SUINF$. Actually for  groups of the order of $N=19$
the profile shows a double maximum structure, instead of the  single maximum
one.  

There are, nevertheless, several features which quite consistently seem  to
become more exact as $N \rightarrow \infty$.
One of the most remarkable is that 
gauge invariant quantities tend to be constant in space~\cite{new}. 
The final conclusions of this analysis are currently under study
(Ref.\cite{gpg-am}).

\subsection{Other results on classical solutions}
In this section,  we would like to comment   about the result
obtained  by P.~van~Baal and P.~Braam~\cite{pvbpb} concerning the
non-existence of $Q=1$ instanton solutions on the torus without twist (See
also \cite{schenk}).
A  recent  less technical discussion of this result is done in
Ref.~\cite{pvb5}. The result is based on Nahm's transformation~\cite{nahm,c+g}.
This transformation takes an  $\UN$ self-dual gauge
field  $\Amu$  on the torus  with topological charge $Q$, and gives rise to a
new self-dual  $U(Q)$ gauge field living in the dual torus, and whose
topological charge is $N$. If we start with an  $\SUN$ gauge field, the
transform is an $SU(Q)$ gauge field. As a consequence,  it is clear that
there cannot exist  an $\SUN$ self-dual gauge field with unit topological charge,
since its Nahm's transform would make no sense.

For completeness, let us  sketch the definition of the Nahm's transform for the gauge
potential  $\Amu$. First,  one  introduces
a whole family of gauge inequivalent self-dual gauge fields as follows:
\be 
 {\bf A}_{\mu}(x,z) = {\bf A}_{\mu}(x) + 2 \pi z_{\mu} {\bf I}\ , 
\ee
where $0 \le z_{\mu} < 1/l_{\mu}$  are real numbers. They label the points
of the dual torus of size $\frac{1}{l_0} \times \ldots \times \frac{1}{l_3}$.
This is so, because there exist  a gauge transformation that maps ${\bf
A}_{\mu}(x,z_{\nu})$ into ${\bf A}_{\mu}(x,z_{\nu}+1/l_{\nu})$.
The next step  is to consider the   Dirac operators corresponding to this
family of gauge potentials. It is crucial that these operators have exactly $Q$
positive chirallity ($\psi^{(i)}(x,z)$ for i=1 \ldots Q) and no negative 
chirallity solutions (the difference is
fixed by the index theorem). Then the Nahm's transform is given by:
\be
\hat{A}_{\mu}^{i j}(z)= \imath\, \int\, dx\  \psi^{\dagger\, (i)}(x,z)
\frac{\partial}{\partial z_{\mu}}
\psi^{(j)}(x,z)\ . 
\ee
The proof  that $\hat{A}_{\mu}^{i j}(z)$ is self-dual, follows very closely
the ADHM construction \cite{adhm,christ}. 

It is  tempting to think that the Nahm's transform could allow the
construction of new self-dual solutions, starting  from  known ones.
However, it is known that the Nahm's transform of a constant field 
strength solution is  another constant field strength solution. 
An explicit example is given in Ref.\cite{pvb5}.


\section{Fluctuations around classical solutions}

In this section we will very briefly comment a few things about fluctuations 
around classical solutions. The first point we want to mention concerns 
zero modes. If we have a classical solution and we consider a small
perturbation around it,
the variation of the action  is to  first order equal to zero. To second
order, we have a quadratic form given by an 
operator whose spectrum is the spectrum of fluctuations. For a minimum action 
solution all eigenvalues are positive or zero. The corresponding
zero-eigenvalue
eigenvectors are called  {\em zero-modes}. 
If the classical solution depends on several parameters,  the
variation with respect to each parameter gives rise to a zero-mode. Hence,
there are at least as many independent zero modes as parameters of which the solution 
depends.  For a gauge theory,  there are always  zero-modes  present, associated with gauge
transformations of the solution. Hence, what  one is  really interested in, 
is in those zero-modes, giving rise to parameters,  which are not associated to gauge transformations.
The space of these parameters is what is called the moduli space. The number
of independent zero-modes not associated with gauge transformations gives
its dimensionality.        

For  the case of self-dual  solutions, it was found some time
ago~\cite{schwartz,atiyah}, that    the Atiyah-Singer index theorem can tell us the
dimensionality of the moduli space $\mathbf{d}$, even if we do not know the analytic
expression of the solutions.
For  ordinary instantons this is well-known: $\SUT$ self-dual solutions
on the sphere are known to depend on $\mathbf{d}=8 |Q| -3$ parameters ($Q$ is the
topological charge). This was known
before the ADHM method  gave us a recipe to construct them. However,
the result depends both on the gauge group and on the base-manifold (See
Refs~\cite{atiyah2,bernard}).
In particular, for the torus $\TF$ the result is $\mathbf{d}= 4\, c_A(G)\, |Q|
$, where $c_A(G)$ is the quadratic Casimir in the adjoint representation of
the group G. Hence, $\SUN$ self-dual solutions depend on $\mathbf{d}=4\, N\,
|Q|$ parameters. 

As a consequence, of the previous formula, the $Q=\frac{1}{N}$ instantons
which appear in twisted $\SUN$ theories, depend on 4 parameters. These
parameters are associated to the translational zero-modes in 4 dimensional
space. There is no zero-mode associated with scale invariance, as for BPST
instantons, since this symmetry is broken by the torus size.
Actually, the form of  $\mathbf{d}$ is consistent with the interpretation of
the space of  self-dual  solutions, as a liquid of $Q=\frac{1}{N}$ instantons
done in Ref.~\cite{Investigating}. 

An analysis of the  fluctuation spectrum  around 
the constant field-strength solutions of subsection 7.2 has been
carried by P.~van~Baal~\cite{vanbaal6}. He concluded that the only  solutions
which are stable are the self-dual (or anti-self-dual) ones. 
The eigenfunctions are given in terms of generalized Riemann theta functions.

\section{Concluding remarks}
In the previous pages, we have reviewed some of the most important results
obtained for classical \YM\ fields on the torus, many of which turn out
to be relevant when one studies the quantum mechanical system. The second
part of my lectures  at  Pe\~niscola, covered this aspect. However, 
the  written-up  version of this part is not yet ready and will appear 
later.  Unfortunately, 
a good deal of our motivation for plunging into the study of the classical
system lies  precisely in the second part.  We hope this will not discourage
the interested reader. 

To conclude, I will just mention  that some topics have been left out  due
to lack of space and time. For example, we have not included the results  obtained
about sphalerons in recent years\cite{sphal1,sphal2}.  In any case, I hope that the present review 
will turn to be useful for those interested in studying the subject. 

\end{document}